\documentclass[a4paper,fleqn,usenatbib]{mnras}

\usepackage{newtxtext,newtxmath}

\usepackage[T1]{fontenc}
\usepackage{ae,aecompl}

\usepackage{graphicx}	
\usepackage{amsmath}	
\usepackage{amssymb}	

\title[The Extreme Faint End of the UV Galaxy Luminosity Function at $z \sim 6$]{The Extreme Faint End of the UV Luminosity Function at $z\sim6$  Through Gravitational Telescopes\thanks{Based on observations made with the NASA/ESA Hubble Space Telescope, which is operated by the Association of Universities for Research in Astronomy, Inc., under NASA contract NAS 5-26555. These observations are associated with programs 13495, 11386, 13389, and 11689. STScI is operated by the Association of Universities for Research in Astronomy, Inc. under NASA contract NAS 5-26555. . This work utilizes gravitational lensing models produced by PIs Bradac, Natarajan \& Kneib (CATS), Merten \& Zitrin, Sharon, and Williams, and the GLAFIC and Diego groups. This lens modeling was partially funded by the HST Frontier Fields program conducted by STScI. The Hubble Frontier Fields data and the lens models were obtained from the Mikulski Archive for Space Telescopes (MAST)}: a comprehensive assessment of strong lensing uncertainties}

\author[Atek et al.]{
Hakim Atek,$^{1}$\thanks{E-mail: hakim.atek@iap.fr}
Johan Richard,$^{2}$
Jean-Paul Kneib$^{3,4}$
and Daniel Schaerer$^{5}$
\\
$^{1}$Institut d'astrophysique de Paris, CNRS UMR7095, Sorbonne Universit\'e, 98bis Boulevard Arago, F-75014 Paris, France  \\
$^{2}$Univ Lyon, Univ Lyon1, Ens de Lyon, CNRS, Centre de Recherche Astrophysique de Lyon UMR5574, F-69230, Saint-Genis-Laval, France \\
$^{3}$Laboratoire d'Astrophysique, Ecole Polytechnique F\'ed\'erale de Lausanne, Observatoire de Sauverny, CH-1290 Versoix, Switzerland \\
$^{4}$Aix Marseille Universit\'e, CNRS, LAM (Laboratoire d'Astrophysique de Marseille) UMR 7326, 13388, Marseille, France \\
$^{5}$Observatoire de Gen\`eve, Universit\'e de Gen\`eve, 51 Ch. des Maillettes, 1290, Versoix, Switzerland \\
}

\date{Accepted XXX. Received YYY; in original form ZZZ}

\pubyear{2017}

\begin{document}
\label{firstpage}
\pagerange{\pageref{firstpage}--\pageref{lastpage}}
\maketitle

\begin{abstract}
With the Hubble Frontier Fields program, gravitational lensing has provided a powerful way to extend the study of the ultraviolet luminosity function (LF) of galaxies at $z \sim 6$ down to unprecedented magnitude limits. At the same time, significant discrepancies between different studies were found at the very faint end of the LF. In an attempt to understand such disagreements, we present a comprehensive assessment of the uncertainties associated with the lensing models and the size distribution of galaxies. We use end-to-end simulations from the source plane to the final LF that account for all lensing effects and systematic uncertainties by comparing several mass models. In addition to the size distribution, the choice of lens model leads to large differences at magnitudes fainter than $M_{UV} = -15~$ AB mag, where the magnification factor becomes highly uncertain. We perform MCMC simulations that include all these uncertainties at the individual galaxy level to compute the final LF, allowing, in particular, a crossover between magnitude bins. The best LF fit, using a modified Schechter function that allows for a turnover at faint magnitudes, gives a faint-end slope of  $\alpha = -2.01_{-0.14}^{+0.12}$, a curvature parameter of $\beta = 0.48_{-0.25}^{+0.49}$, and a turnover magnitude of $M_{T} = -14.93_{-0.52}^{+0.61}$. Most importantly our procedure shows that robust constraints on the LF at magnitudes fainter than $M_{UV} = -15~$ AB remain unrealistic. More accurate lens modeling and future observations of lensing clusters with the {\em James Webb Space Telescope} can reliably extend the UV LF to fainter magnitudes.

\end{abstract}




\section{Introduction} 

The epoch of cosmic reionisation around $z \sim 6$ to 10 \citep[e.g.,][]{fan06, planck16} has seen the neutral hydrogen content of the Universe become ionised, ending the period of dark ages. The identification of sources responsible for such dramatic phase transition remains a major question in extragalactic astronomy. Early star-forming galaxies could well be the best candidates to drive reionization \citep{robertson13,finkelstein15,bouwens15} as the study of their luminosity function at $z > 6$ reveals a steep faint-end slope that indicates a significant contribution from faint galaxies \citep{bunker10,oesch10,bouwens11,mclure13}. Much of the progress in the field was achieved in deep blank fields observed with the {\em Hublle Space Telescope} ({\em HST}), which now reaches an observational limit around an absolute magnitude of M$_{\rm UV} \sim -17~$AB. The total UV luminosity density emitted by galaxies down to such limit is insufficient to drive reionisation and an extrapolation of the LF is used to account for the contribution of faint galaxies. However, the abundance of this population of faint galaxies strongly depends on the faint-end slope and the magnitude cutoff of the UV LF, which are uncertain. 

Massive galaxy clusters act like cosmic telescopes, magnifying background galaxies in the strong gravitational lensing regime, and as such offer a viable route to reach beyond the current observational limit \citep{maizy10,kneib11,sharon12,postman12,richard14a,coe15}. The Hubble Frontier Fields (HFF) program has delivered the deepest observations of lensing clusters to date, with 840 {\em HST} orbits reaching $\sim 29$ AB mag limit in seven optical and near-IR bands \citep{lotz17}, in addition to supporting space {{\em Spitzer, Chandra}} and ground-based ({\em VLT, ALMA}) observations. As part of the HFF project, several groups produced mass models for each cluster, which were made public in order for the community to interpret high-redshift observations \citep[e.g.,][]{jauzac14,johnson14,grillo15,ishigaki15,hoag16,mahler18}. 

The first studies of $z > 6$ UV luminosity functions relying on HFF data and these lensing models already discovered the faintest high-redshift galaxies \citep{atek14b,zheng14,yue14,ishigaki15}, extending the LF by two magnitudes. Combining more lensing clusters, later studies pushed the detection limits down to M$_{\rm UV} \sim -13~$AB \citep[][hereafter, I18 and B17, respectively]{kawamata16,castellano16,laporte16,ishigaki18,bouwens17b} and even to M$_{\rm UV} \sim -12~$AB \citep[][hereafter, L17]{livermore17}. While luminosity functions from different groups tend to agree well on the global shape of the LF at  $M_{\rm UV} < -17$ mag, significant discrepancies appear at the very faint end, a territory where only sources with high magnification can be selected. For example, the most recent work by L17 shows significantly higher values of the LF at $M_{\rm UV} > -17$ compared to results of B17, and I18 find a steeper faint-end slope compared to L17 and B17. While B17 find signs of a turnover in the LF at $M_{\rm UV} > -15$ and a faint-end slope of $\alpha = - 1.91 \pm 0.04$, L17 report a steep faint-end slope of $\alpha -2.10 \pm 0.03$ and strong evidence against a possible turnover at $M_{\rm UV} < -12.5$. This example highlights the challenges encountered by this type of studies to constrain the extreme faint-end of the UV LF at $z\sim6$. The observed discrepancies in those studies might result from the differences in dropout catalog selection, the size distribution adopted in the completeness simulations to compute the survey volumes (B17), and also from the different approaches in the treatment of lensing uncertainties.

In this paper we present for the first time a comprehensive assessment of the lensing uncertainties affecting the UV LF using end-to-end simulations from the source plane to the final UV LF. The focus of this study is the faint-end (beyond $M_{\rm UV} = -15$) of the LF and how the choice of a given size distribution and a lensing model can impact the result. The paper is organized as follows. In Section \ref{sec:obs} we describe the full HFF imaging dataset used in the study and in Section \ref{sec:sample} we present the selected sample of $z \sim 6$ galaxies. The lensing models are described in Section \ref{sec:models}. Our end-to-end simulation procedure, used to both compute the effective survey volume (Section \ref{sec:volume}) and quantify the uncertainties is detailed in Section \ref{sec:simulations}. The final UV LF and the associated uncertainties are discussed in Section \ref{sec:LF}, while the UV luminosity density is computed in Section \ref{sec:density} before the conclusion given in Section \ref{sec:conclusion}. Throughout the paper we adopt standard cosmological parameters: $H_0=71$\ km s$^{-1}$\ Mpc$^{-1}$, $\Omega_{\Lambda}=0.73,$\ and $\Omega_{m}=0.27$. All magnitudes are expressed in the AB system \citep{oke83}.

\vspace{1cm}

\section{HFF Observations}
\label{sec:obs}

The Frontier Field program obtained deep {\em HST} imaging data in the optical and near-infrared for a total of six lensing clusters and flanking fields. The Advanced Camera for survey (ACS) was used for the optical coverage with three broad-band filters F435W, F606W, and F814W. The Wide Field Camera Three (WFC3) was used for the NIR observations with four filters F105W, F125W, F140W, and F160W. HST obtains WFC3/IR and ACS optical simultaneously for each pair cluster-parallel field in the first epoch before switching instrument positions in the second epoch. Observations took place between 2013 and 2016 over HST cycles 21 to 23 devoting 140 orbits to each cluster/parallel pair. In table \ref{tab:obs} we give the limiting magnitude reached in each filter and measured in a circular aperture with 0.4\arcsec diameter (cf. Table \ref{tab:obs}).

In the present analysis we use the high level science products made available by the HFF data reduction team at the Space Telescope Science Institute\footnote{\url{http://www.stsci.edu/hst/campaigns/frontier-fields/}} (STScI). Data products include drizzled images, and weight maps from the FF program but also ancillary data from other observing programs. We used the ACS mosaics generated using the "self-calibration" method and the WFC3/IR mosaics that were corrected for time-variable sky background. The detailed reduction procedure and science products can be found at \url{https://archive.stsci.edu/prepds/frontier/}.

All the images were first matched to the same PSF with the F160W frame as reference. We created deep image stacks using a weighted combination of the four filters in the IR and the two bluest filters in the optical for the detection and non detection, respectively, criteria for $z \sim 6$ dropout selection. For $z \sim 8$ galaxies the combined images include the three reddest IR filters for the detection and the three ACS filters for the non detection criteria. The cluster fields are affected by intra cluster light (ICL) and the bright cluster galaxies (BCGs) that significantly reduce our ability to detect galaxies around the cluster core. In order to mitigate these effects and improve the effective depth of the cluster fields we performed a median filtering procedure to subtract the foreground light. We used a median filter of 2\arcsec $\times$ 2\arcsec for all individual and stacked images. More sophisticated techniques have been used to correct for the cluster light contamination such as wavelet decomposition (L17), GALFIT \citep{peng02} combined with median filtering (B17), and other iterative modeling techniques of the bright cluster galaxies \citep{shipley18}.

\begin{table*}
\caption{Magnitude limits of {\em HST} observations in the six HFF clusters. The depth of the images are 3-$\sigma$ magnitude limits measured in a 0.4\arcsec\ circular aperture.}
\label{tab:obs} 
\begin{tabular}{lccccccccc}
\hline

Field & RA & DEC & & ACS &  & & WFC3  & &   \\ 
 & J2000 & J2000 & F435W& F606W& F814W  & F105W& F125W  & F140W &  F160W  \\ \hline

A2744  &00:14:21.2& $-$30:23:50.1 & 28.8 & 29.4 & 29.4 & 28.6 & 28.6 & 29.1 & 28.3 \\
MACS0416 &04:16:08.9 & $-$24:04:28.7 & 30.1 & 29.1 & 29.2 & 29.2 & 28.8 & 28.8 & 29.1   \\
MACS0717 & 07:17:34.0 & $+$37:44:49.0 & 29.5 & 28.6 & 29.3 & 28.9 & 28.6 & 28.5 & 28.8 \\
MACS1149 & 11:49:36.3 & $+$22:23:58.1& 28.6 & 28.6 & 28.6 & 28.9 & 29.3 & 29.2 & 30.1 \\
AS1063 & 22:48:44.4 & $-$44:31:48.5 & 30.1 & 29.1 & 29.3 & 29.0 & 28.7 & 28.5 & 28.8 \\
A370 & 02:39:52.9 & $-$01:34:36.5 & 30.1 & 29.1 & 29.3 & 29.0 & 28.7 & 28.5 & 28.4 \\ \hline

\end{tabular}
\end{table*}

\section{The $z \sim$ 6 Galaxy Sample}
\label{sec:sample}

We perform the source extraction using the SExtractor software \citep{bertin96}. In the cluster fields, the ICL-corrected images are used as a detection image while the photometry is performed in the original ones. The isophotal magnitudes {\tt MAG\_ISO} are used for the color-color selection whereas {\tt MAG\_AUTO} values are adopted in calculating the total magnitudes. We adopt Lyman break selection criteria \citep[e.g.,][]{steidel96} similar to those in previous studies of dropout galaxies at these redshifts \citep{bouwens14,finkelstein14}, and which were explained in detail in \citet{atek15a}: 
\begin{align}
\label{eq:criteria7}
 (I_{814} {-} Y_{105})   &>  1.0  \notag \\
(I_{814} {-} Y_{105})   &>  0.6 + 2.0 (Y_{105} {-} J_{125})\\
(Y_{105} {-} J_{125})  &< 0.8 \notag
\end{align}
Galaxies in the sample must also satisfy a detection significance above 5$\sigma$ in at least two IR filters and 6.5$\sigma$ in the IR stacked image. We also require a non detection in both F435W and F606W filters and their stack. This color selection includes both $z \sim 6$ and $\sim 7$ galaxies and is chosen to exclude low-redshift sources that have similar colors to those of high-$z$ galaxies. The most important sources of contamination consists of dust-obscured and evolved galaxies at lower redshift with extremely red colors, and low-mass stars. In order to mitigate this contamination, we investigated the evolution of low-redshift elliptical galaxy templates from \citet{coleman80} and starburst galaxy templates from \citet{kinney96} in the color-color diagram as a function of redshift and extinction. We also simulated the color-track of stars from \citet{chabrier00} templates. An additional visual inspection is performed to rule out spurious detections and point-like sources. A more detailed review of all the potential interlopers is given in \citet{atek15a}. As a natural consequence of strong lensing, we also expect that some of the background sources will have multiple images that need to be accounted for in the final galaxy number counts. The details of this procedure are given in Section \ref{sec:observed}. The final sample contains a total of 300 galaxies in all the cluster fields, with magnification factors ranging from $\mu \sim 1$ to few hundreds, including all models used here.

\section{Lensing Models}
\label{sec:models}
Computing an accurate luminosity function at $z=6-7$ to the faintest luminosities requires a good knowledge of the lensing power of the HFF clusters. Lensing models are essential not only to estimate the magnification of the sources but also the effective survey volume and the completeness function (cf. Section \ref{sec:volume}). For the six clusters we adopt the publicly available lensing models constructed by the CATS team (clusters as telescopes) using version 7.0 of the Lenstool\footnote{\url{http://projets.lam.fr/projects/lenstool/wiki}} software, which follows a parametric approach to map the mass distribution in the cluster \citep{kneib93,jullo07,jullo09}. The strong lensing analysis of the six HFF clusters using the full depth of {\em HST} observations is now complete. The mass reconstructions of the first three clusters were published in \citet{jauzac14, jauzac15, limousin16, lagattuta17}. All the lens models are made publicly available through the public Frontier Fields repository\footnote{\url{https://archive.stsci.edu/prepds/frontier/lensmodels/}}, which includes mass models submitted by several teams: the GLAFIC team \citep{oguri10,ishigaki15}, Bradac et al. \citep{bradac05, hoag16}, Merten \& Zitrin \citep{merten11, zitrin09,zitrin13}, Sharon \& Johnson \citep{johnson14}, and \citet{diego15}. We note that at the time of conducting te present analysis, several teams have not updated their mass reconstruction models and do not include the full HFF observations. The impact of model uncertainties and systematic differences between the models on the computed UV LF are assessed in Section \ref{sec:errors}.

\begin{figure}
   \centering
   \includegraphics[width=9cm]{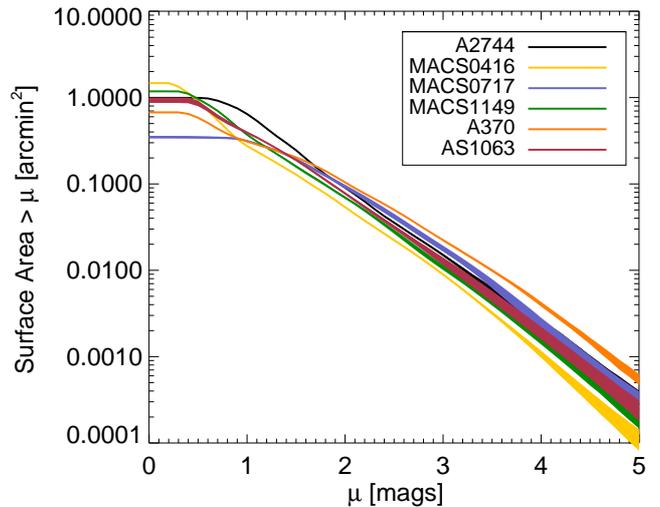} 
   \caption{Cumulative surface area as a function of magnification. The area for each cluster is derived using the CATS model and each colored area is indicative of the model uncertainties following the color code in the legend. The magnification factor $\mu$ is given in units of magnitude}
   \label{fig:surface}
\end{figure}

\begin{figure*}
   \centering
   \vspace{-2cm}
   \includegraphics[width=16cm]{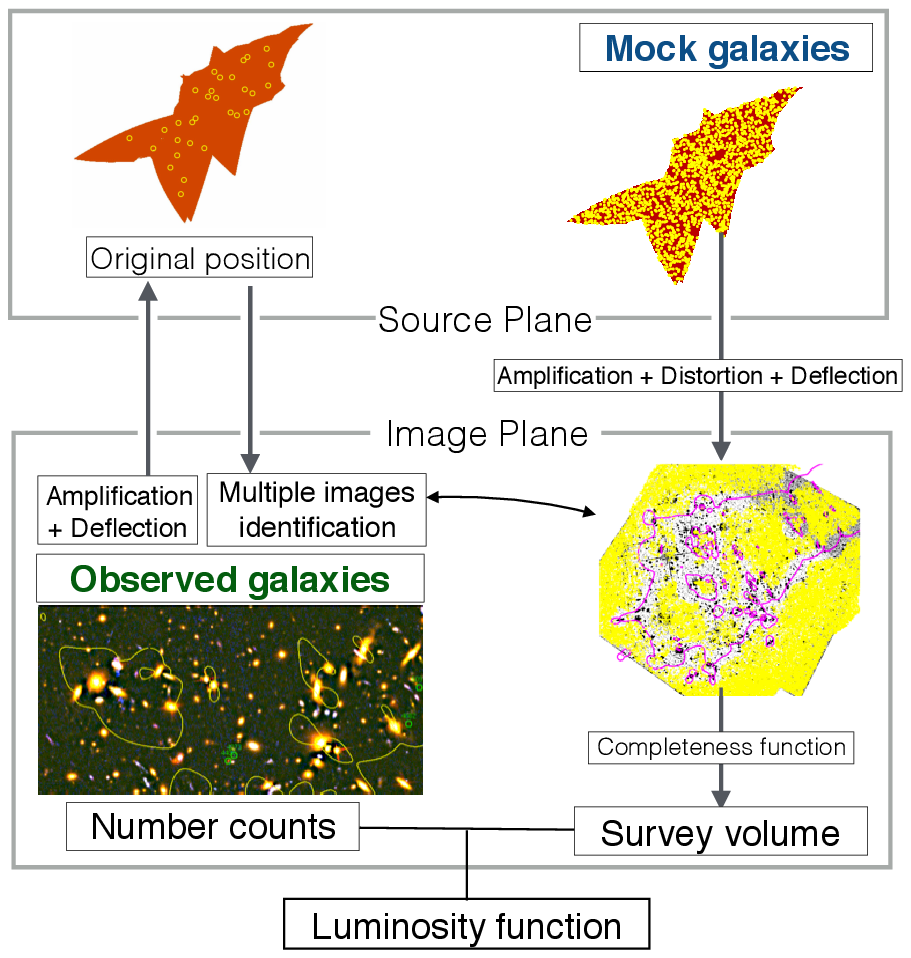} 
   \vspace{-5cm}
   \caption{Illustration of the source plane method in computing the UV luminosity function and assessing the impact of lensing model uncertainties. Magnification maps provide the amplification of the dropout galaxies identified in the {\em HST} images, which are de-lensed into the source plane using the deflection maps and then projected back into the image plane to identify multiple-image systems. In parallel, we create in the source plane a uniform distribution of simulated sources, which are lensed to the image plane using the same mass model as the actual observations. The process includes the amplification and shape distortion of the galaxies and it automatically accounts for multiple images. Here we show directly the real WFC3 image (of MACS0717 in this example) where the galaxies are implemented. Note the rarity of sources in the region of highest magnification around the critical line at $z \sim 6$ shown by the magenta curve. The completeness function computed from the mock galaxies is combined with the reduced surface area to determine the survey volume, and ultimately the luminosity function.} 
   \label{fig:source_plane}
   \label{fig:flow}
\end{figure*}

\section{ end-to-end simulation procedure}
\label{sec:simulations}

Alongside the galaxy number counts and their intrinsic luminosities, we need to estimate the effective survey volume in each cluster. Following \citet{atek14b} we performed all the completeness estimates directly in the source plane. This approach has the advantage of naturally accounting for all the lensing effects that affect the galaxy properties (shape, luminosity) and the survey area. In Sections \ref{sec:err-size} and \ref{sec:err-models}, we use the same technique in our end-to-end simulations to assess the reliability of the lensing models and size distribution in recovering the true UV luminosity function. The flow chart of Fig. \ref{fig:flow} illustrates the different steps throughout the UV LF determination procedure.

\subsection{Observed galaxies}
\label{sec:observed}
The first step is the selection of dropout galaxies in the actual {\em HST} images (cf. Section \ref{sec:sample}) before de-lensing them to the source plane and determine their amplification factor. Then the sources are lensed back to the image plane where we predict the position of the counter-images of each source in the multiple-image area. then around this position we search for dropout galaxies that have a similar redshift probability, color and morphological symmetries in the case of resolved sources. We derive the intrinsic absolute magnitude based on the observed magnitude and the amplification factor. The galaxy redshifts are determined from the peak probability of the photometric redshifts using a modified version of the Hyperz code \citep{bolzonella00,schaerer09}, which includes nebular continuum and emission lines contribution. Finally, the galaxy number counts are computed in magnitudes bins with a bin size of $\Delta$mag =0.5 and corrected for te multiple images. As we will see in the assessment of lensing uncertainties of Section \ref{sec:err-models}, we can use any lensing model to compute the intrinsic magnitudes and identify the multiple images.

\subsection{Simulated galaxies}
\label{sec:simulated}

The second step consists of end-to-end simulations to determine the incompleteness function and the associated uncertainties. For each field, we simulated a set of 10,000 galaxies with randomly distributed redshifts and intrinsic absolute magnitudes and two different light profiles: an exponential disk and de Vaucouleurs profile \citep{ferguson04,hathi08}. The input galaxy sizes follow a log-normal distribution with a mean half light radius of 0.15 \arcsec\ and a dispersion of 0.05 \arcsec\   \citep{bouwens04, hathi08, grazian12, huang13, ono13}. However, the sizes are also luminosity-dependent following recent results on size-luminosity relation of high-redshift galaxies \citet{oesch14, mosleh12, kawamata15}. We discuss in further detail the influence of the size distribution in Section \ref{sec:err-size}. 

The simulated galaxies are randomly distributed directly in the source plane of each cluster, which was reconstructed from the WFC3 field of view using the lens model. The total survey area in the source plane and its evolution as a function of magnification is shown in Figure \ref{fig:surface}. We clearly see that at very high magnifications ($\mu > 5$ mag) the survey area becomes so small that the probability to detect any galaxy is close to zero. A direct consequence of such effect is that the surface density of simulated galaxies in the image plane decreases rapidly moving closer to the critical line, as the amplification increases rapidly as well (cf. Figure \ref{fig:source_plane}). For this procedure, other studies have adopted various methods. I18 adopted a similar approach to ours, where sources are simulated in the source plane using the GLAFIC software. B17 also use a source distribution in the source plane relying only on the magnification map (i.e. without using lensing deflection). An image plane method was adopted in L17, choosing to under-sample the region of high magnification in the source plane. Our source plane approach with a non-uniform distribution of sources in the image plane is supposed to reflect reality since the distribution is uniform in the ``physical'' plane. This method also ensures that galaxy shapes are distorted according to the shear potential, which directly affects the detection limit, hence the completeness function. Most importantly, the completeness function depends on several parameters, some of which are interdependent, such as the magnification and the source position, and need to be simultaneously accounted for.

 Using {\tt Lenstool}, we lensed all the sources towards the image plane. Then we randomly assigned a starburst spectral energy distribution to galaxies among a library of stellar population models from \citet{bc03}. The spectral templates are redshifted and normalized to the observed magnitude, including magnification, in the F125W filter corresponding to the rest-frame UV, while the magnitude in the rest of the filters are computed using their throughput curves. The mock galaxies are added directly to the real HFF images (10 galaxies per image at a time) of each filter and convolved with the PSF of the F160W image. Finally, we follow the same procedure used for the observations to extract the sources and select dropout galaxies at $z \sim 6$. 
 
 \section{Effective Survey Volume}
 \label{sec:volume}
 
 The completeness function is computed by comparing the output catalog with the original input one as a function of the intrinsic magnitude. The effective survey volume is then given by:

 \begin{eqnarray}
 \label{eq:volume}
V_{eff} = \int_{0}^{\infty}  \int_{\mu > \mu_{min}} \frac{dV_{com}}{dz}~ f(z,m,\mu) ~d\Omega(\mu,z) ~dz
\end{eqnarray}

where $V_{com}$ is the comoving volume, $f$ the completeness function, which depends on the redshift, the apparent magnitude $m$ and the magnification factor $\mu$. $d\Omega(\mu,z)$ is the surface element corresponding to a magnification $\mu$ at a given redshift $z$ and $\mu_{min}$ is the minimal magnification value at which a galaxy with a magnitude $m$ can be detected.

\begin{figure}
   \centering
   \includegraphics[width=8.5cm]{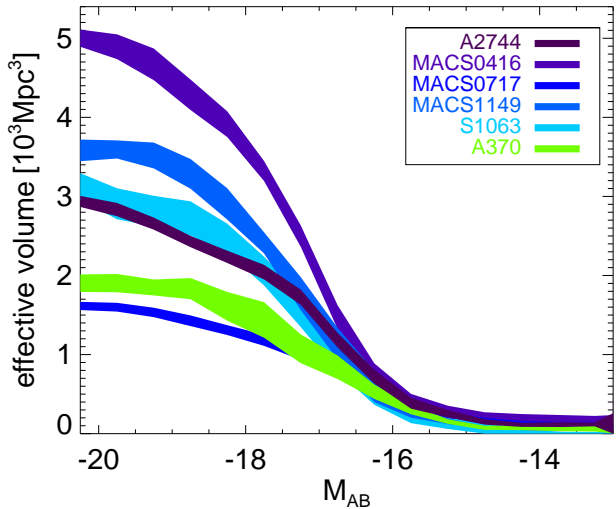} 
   \caption{The effective survey volume as a function of the intrinsic absolute magnitude in the UV. Each curve shows the result for an individual cluster. The volume is computed from the completeness function and the surface area of each cluster following equation \ref{eq:volume} and using the CATS lens models. The colored areas around the curves represent 1$-\sigma$ uncertainties. }
   \label{fig:volume}
\end{figure}

The resulting effective survey volume for each cluster field marginalized over the intrinsic absolute magnitude is shown in Figure \ref{fig:volume}. The maximum volume depends mainly on the total surface area with no magnification (cf. Fig. \ref{fig:surface}) in each cluster, the maximum completeness being around 80\%. The curves drop rather quickly at the bright end before flattening out at the faint-end, with the help of lensing magnification. Some clusters are better lenses than others in detecting the faintest galaxies, which is a direct consequence of the balance between the magnification power and the reduction of survey area, but also the lens geometry.

\begin{figure*}
   \centering
    \includegraphics[width=8.5cm]{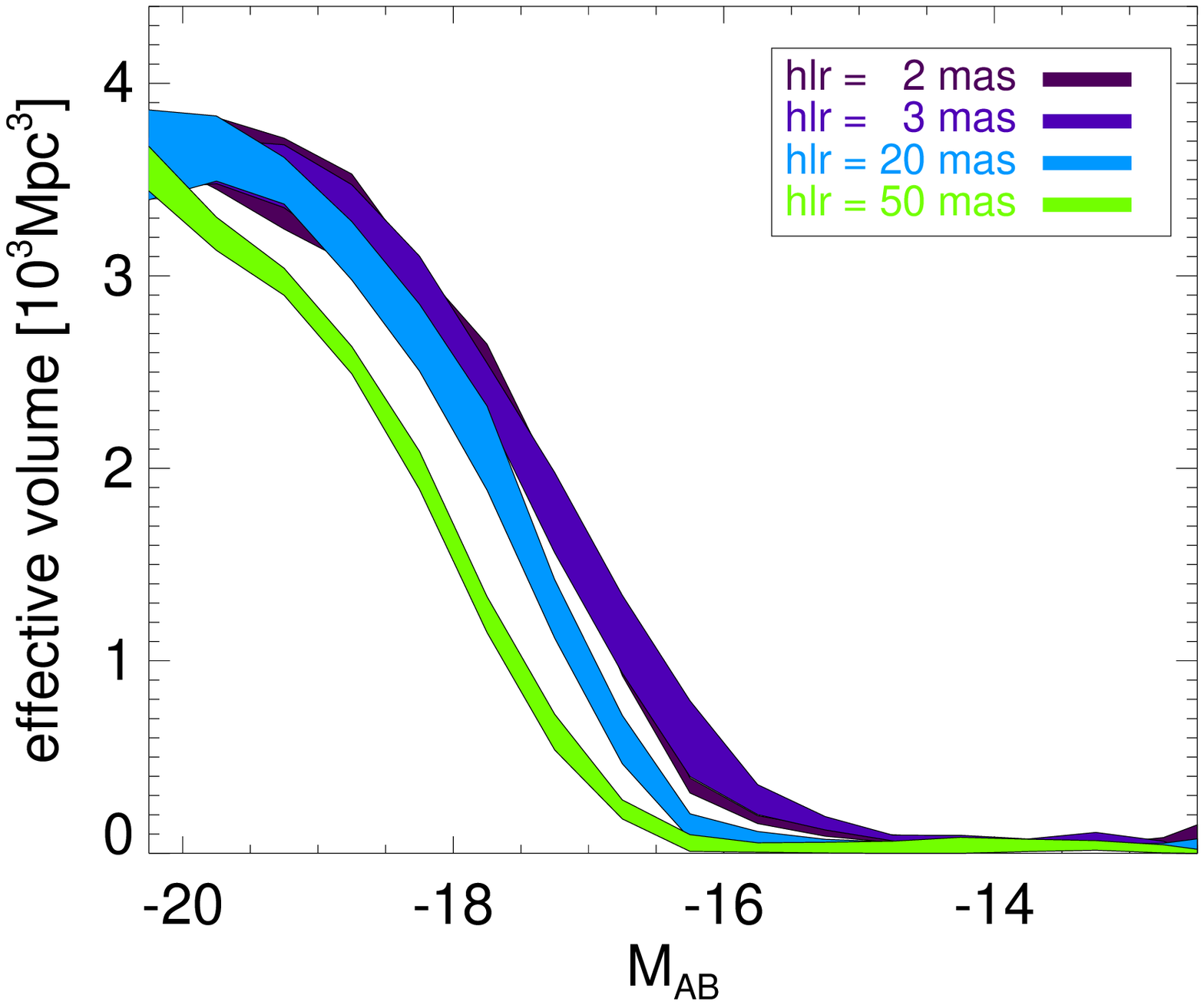} 
       \includegraphics[width=8.5cm]{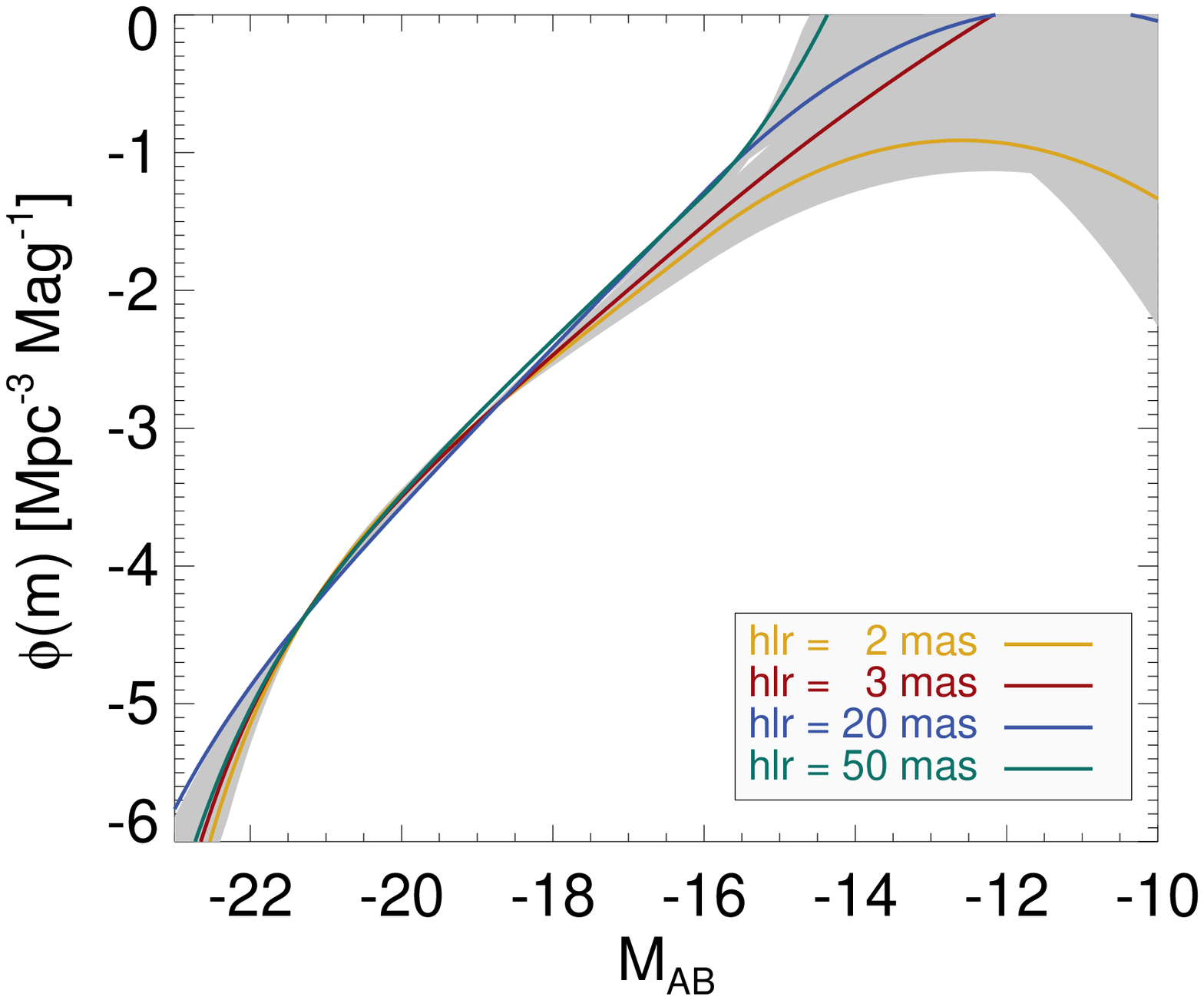}
   \caption{Impact of galaxy size distribution on the effective survey volume and the UV LF of MACS1149. In all cases the simulated sizes are based on a log-normal distribution in the source plane that is lensed to the image plane. The legend indicates te mean half light radius of galaxies fainter than $M_{AB} = -16$ mag in each simulation. The left panel shows the effective survey volume for each size distribution wile the right panel sows its impact on the final UV LF at $z\sim6$. The error regions for each size distribution are represented in grey in the LF plot and overlap with each other.}
   \label{fig:size}
\end{figure*}

\section{The UV Luminosity Function at $z \sim 6$}
\label{sec:LF}

With the effective survey volume in hand, the number counts can now be used to compute the intrinsic UV luminosity function in each individual cluster as follows

\begin{eqnarray}
\phi(M_{i})dM_{i} = \frac{N_{i}}{V_{eff}(M_{i})},
\end{eqnarray}
where $N_{i}$ is the number of galaxies in each magnitude bin and $V_{eff}(M_{i})$ the effective survey volume in the $i$th bin of absolute magnitude $M_{i}$.

Before computing the final UV LF at z$\sim$6 resulting from the combination of all the clusters, we will use our simulation procedure in the source plane to perform, for the first time, a realistic assessment of the different uncertainties that affect the UV LF.

In order to fit the observed LF points, we adopt a modified Schechter function \citep[cf.][]{bouwens17b} to allow for a potential turnover of the LF at the faint end, which is in line with most of the theoretical models that predict a drop in the UV LF due to star formation inefficiency in small dark matter haloes \citep[eg.][]{jaacks13, gnedin16,yue16}. The general Schechter form is given by:

\begin{eqnarray}
\phi(M)=\frac{\rm{ln}(10)}{2.5} \phi^{\star}10^{0.4(\alpha+1)(M^{\star}-M)} exp(-10^{0.4(M^{\star}-M)}), 
\label{eq:schechter}
\end{eqnarray}
 while for magnitudes fainter than $M_{AB} = -16$ we multiply the Schechter function by the turnover term

\begin{eqnarray}
10^{-0.4 \beta(M+16)^{2}},
\label{eq:turn}
\end{eqnarray}
where $\beta$ is the curvature parameter. The UV LF will have a downward turnover for $\beta > 0$, but an upward turnover is also permitted for $\beta < 0$. In the following, in the case of $\beta >0$, we define the turnover magnitude $M_{T}$, wich corresponds to the magnitude for which $(d\phi/dM)_{M = M_{T}} = 0$.   

\begin{figure*}
   \centering
    \vspace{-5cm}
   \includegraphics[width=17cm]{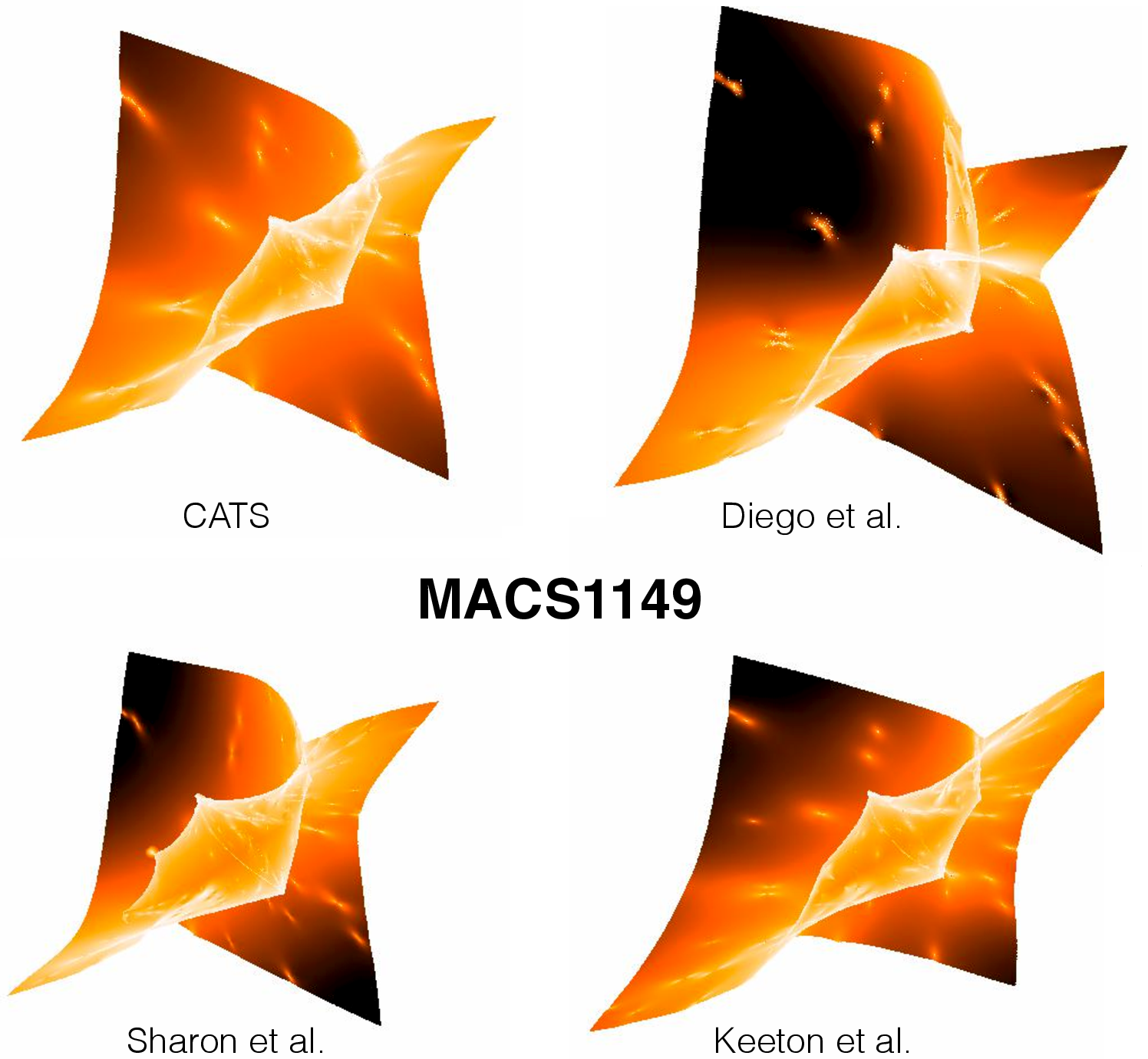} \hspace{-5cm} 
   \vspace{-6cm}
   \caption{Source plane reconstruction for MACS1149 cluster using four different lensing models \citep{richard14, jauzac16,johnson14,ammons14,diego16} for which deflection maps were submitted to the HFF lensing project. The mass models affect not only the magnification maps but also the total surface area in the source plane and the spatial distribution of the sources when lensed back to the image plane.}
   \label{fig:source_planes}
\end{figure*}

\subsection{Impact of Source Size Distribution on the UV LF}
\label{sec:err-size}

Thanks to the sensitivity and the high spatial resolution of the ACS and WFC3 instruments onboard {\em HST}, important progress has been made during the past years in the measurements of sizes of high-redshift galaxies \citep[e.g.,][]{oesch10, huang13, holwerda15}. Besides the importance of these constraints on the early assembly and the structure evolution of galaxies, it directly affects the completeness calculations, hence the derived UV luminosity function. In our simulations presented in Section \ref{sec:volume}), we adopted a log-normal distribution with a mean half-light radius of 150 mas and applied a size-luminosity relation in the form of $r_{hl} \propto L^{0.5}$, derived for lensed galaxies \citep{kawamata15,bouwens17a}. Recent results from lensing fields suggest that the faintest galaxies in the high-redshift samples might have an even smaller size than what has been found in previous studies. In particular, \citet{bouwens17c} find that galaxies down to $M_{AB} \sim -15$ mag have near-point source profiles with typical half-light radii below 30 mas.

Considering the spatial distortion due to lensing shear, the effects of the size distribution might be more significant compared to blank fields. Stacking very faint high-redshift galaxies in HFF observations, \citet{bouwens17a} find that they do not show the apparent size expected for the calculated shear at their location if one assumes the common sizes used in the literature \citep[e.g. the size-luminosity relation of][]{shibuya15}. Rather, the observed sizes are compatible with intrinsic half-light radii around 5 mas. To assess the impact of such size differences on the shape of the UV LF, we adopted three different size distributions for the simulated galaxies in the source plane that are close to what is commonly adopted in the literature. In combination with the size-luminosity relation, galaxies fainter than $M_{AB} \sim -16$ will have hlr=3 mas \citep{bouwens17c}, hlr=20 mas \citep{atek14b,castellano16,ishigaki18}, and hlr=50 mas (L17). The left panel of Fig. \ref{fig:size} shows the effect of such variation in galaxy sizes on the completeness function. Then such completeness functions are applied to the observed number counts in MACS1149 to derive the UV luminosity functions for each size distribution (Fig. \ref{fig:size}, right panel). 

It is clear that adopting smaller sizes for very faint galaxies leads to a larger recovery fraction, hence a larger effective survey volume, which in turn yields a shallower faint-end slope and even a turnover for the smallest size distributions. On the other and, as it can be seen in Fig. \ref{fig:size}, larger simulated galaxies with 50 mas lead to an upward turnover in the LF. This effect might explain the steepening of the LF at the faint-end found by L17, who assume that sources have a median half-light redius of 90 mas at $z \sim 6$. Assuming sizes below 10 mas, \citet{bouwens17b} find a shallower slope as indicated in their Figure 11 because of a higher completeness at the faint-end. Most importantly, our Figure \ref{fig:size} shows that uncertainties on the size distribution and on the extrapolation of the size-luminosity relation beyond $M_{AB} \sim -16$ lead to large uncertainties on the faint-end of UV LF, in particular beyond $M_{AB} \sim -15$, even in the case where extremely small half-light radii (2-3 mas) are adopted. For the final LF determination in Section \ref{sec:errors}, we will use the results obtained for MACS1149 to account for the uncertainties related to the size distribution.

\begin{figure*}
   \centering
   \includegraphics[width=8.7cm]{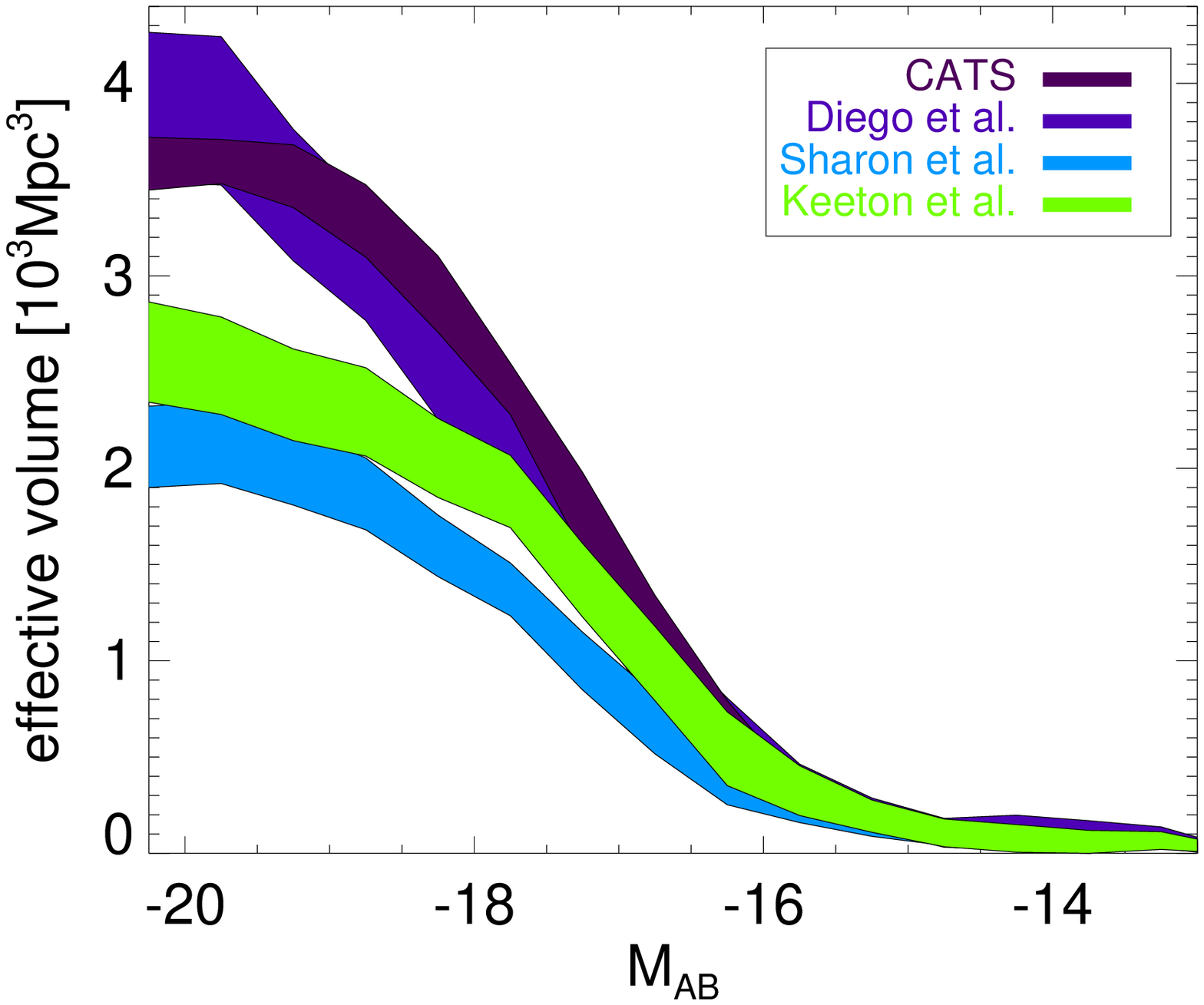} 
   \includegraphics[width=8.8cm]{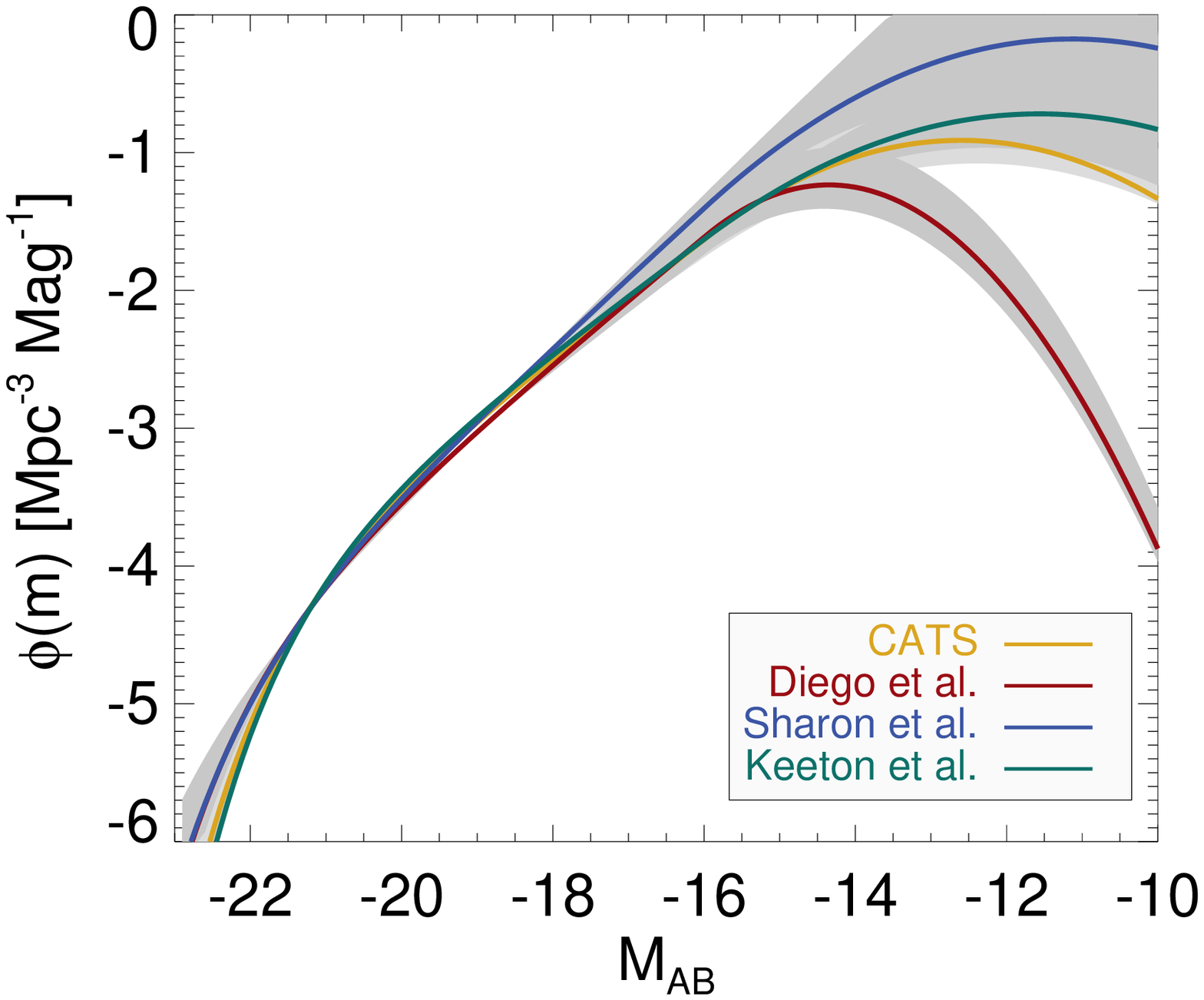} 
   \caption{Impact of cluster mass models on the UV luminosity function in MACS1149: {\em Left:} the effective survey volume (and associated 1$\sigma$ uncertainties) as a function of absolute intrinsic magnitude for each lensing model. The effective volume depends mainly on the geometrical configuration, the magnification map, and the surface area (cf. Fig \ref{fig:source_planes}) of each model. {\em Right:} Derived UV luminosity function using each cluster model and associated uncertainties. All models support a turnover in the faint-end of the LF but at different magnitudes and with different slopes. Beyond an absolute magnitude of M$_{\rm{UV}} =-15$ mag, the differences are too large to robustly constrain the LF shape.}
   \label{fig:models}
\end{figure*}

\subsection{Impact of lensing uncertainties on the UV LF}
\label{sec:err-models}

The present end-to-end simulation procedure allows us to incorporate any lensing model in several steps of the process depicted in Fig. \ref{fig:flow}. The choice of the lensing model will not only affect the amplification factor but also the survey area, the completeness function, and the multiple-images identification and positions. For the first time, the end-to-end nature of our forward modeling handles all these aspects, hence providing realistic estimates of the systematic uncertainties by comparing the results of different lensing models. 

In addition to the CATS models, we selected among the public models three different teams who provided all the necessary information to our simulations: amplification maps, deflection maps, and shear values projected along both directions ($\gamma_1$ and $\gamma_2$). These high resolution maps have been ingested into Lenstool and interpolated to treat them with the same procedures as the CATS models. Based on the published maps from the CATS team, we have tested that this interpolation does not affect our results compared to the Lenstool parametric model. 

The first comparison is the reconstruction of the source plane of the same cluster (the example shown here is MACS1149) using each mass model. Figure \ref{fig:source_planes} highlights large differences in the size of the source plane, therefore the total survey area, where for example the Diego et al. model yields a source area nearly twice as large as that of Sharon et al. Furthermore, the differences in the shape of the source plane also lead to different magnification and different positions for the multiple images when lensed to the image plane. Both actual and simulated images are impacted by the choice of lensing model. Firstly, the adopted model determines the amplification factor of the observed galaxies and the position of the counter-images. Secondly, it determines the amplification, the position, and the distortion of the simulated galaxies (cf. Section \ref{sec:simulations}), which are all linked to each-other. Overall, in combination with the source plane area, the mass model will significantly affect the resulting completeness function.

In Figure \ref{fig:models}, we show the results of the different lensing models on both the effective survey volume as a function of intrinsic absolute magnitude and on the final UV luminosity function of MACS1149. We can see that for the same cluster, the total survey volume can vary by a factor of 2. Differences in intrinsic model uncertainties are also reflected by the dispersion in each curve. The overall shape as a function of intrinsic magnitude also varies from a model to another. The right panel of the same figure shows that the combination of systematic and intrinsic uncertainties lead to large differences in the final UV luminosity function beyond an intrinsic magnitude of M$_{\rm{UV}} =-15$ mag. This is important since it now becomes clear that, with the current depth and lensing models, we will not be able to put robust constraints on the UV LF shape beyond such magnitude limit. It is noteworthy that despite significant differences in the survey volume between the models, the luminosity functions agree well for magnitudes brighter than M$_{\rm{UV}} =-16$ mag. This is essentially due to the fact that for relatively small amplification factors, the variations of the amplification and the effective volume are anti-correlated and tend to maintain the same faint-end slope. Furthermore, relative errors on the effective volume are smaller than what is observed at M$_{\rm{UV}} >-16$ mag.

In the next section we use the same procedure in each cluster to assess the lensing uncertainties in the final LF which combines the six clusters.

\subsection{Final constraints on the UV LF}
\label{sec:errors}

\begin{figure*}
   \centering
   \includegraphics[width=8.8cm]{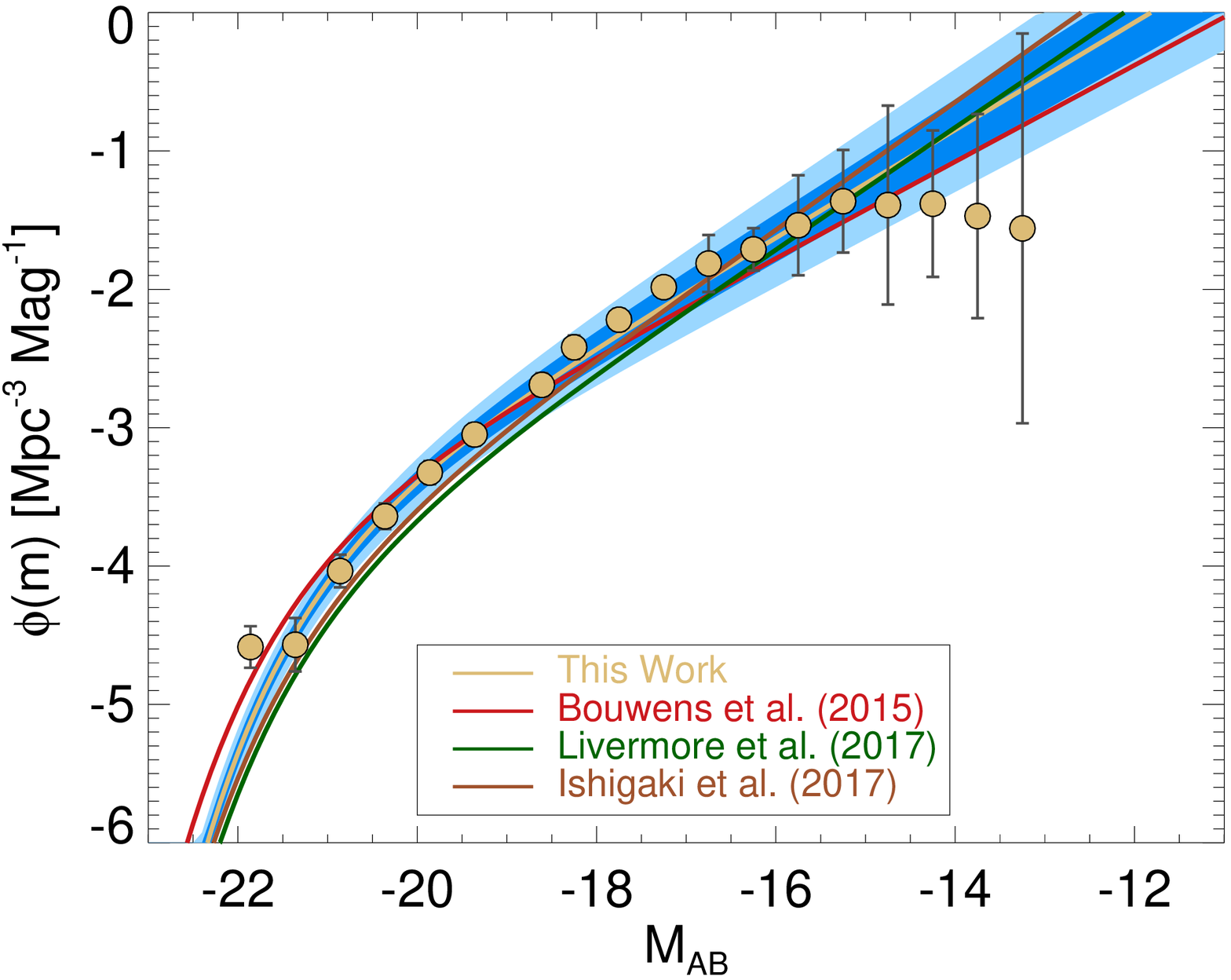}  
     \includegraphics[width=8.8cm]{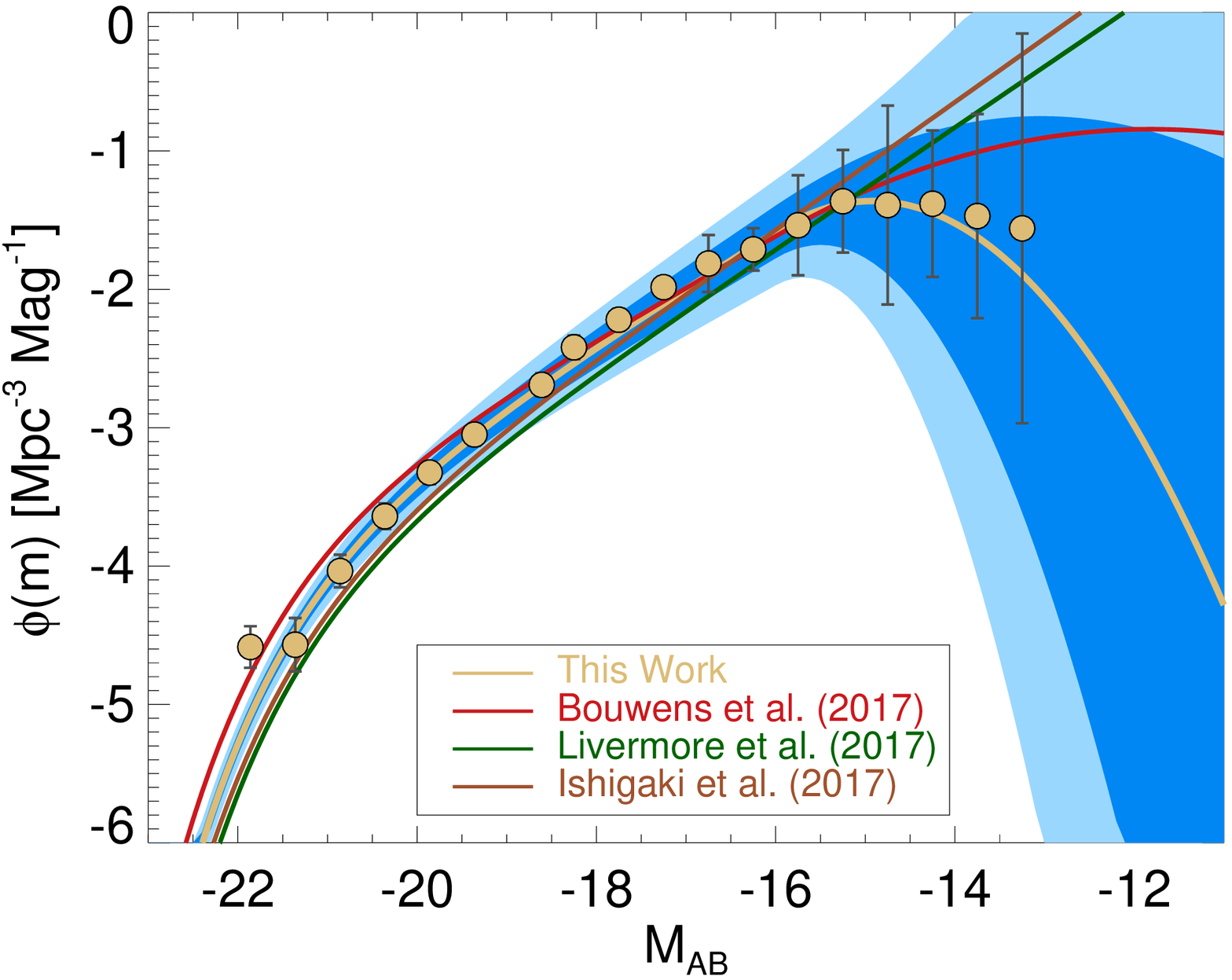} 
   \caption{The final UV luminosity function at $z \sim 6$. The {\bf left panel} shows the best fit of the UV luminosity function using a Schechter form (gold curve) with the associated 1-$\sigma$ (2-$\sigma$) uncertainties in dark (light) blue areas. Measurements (gold data points with 1-$\sigma$ errors) at the very faint-end clearly depart from a simple Schechter form. Therefore, the fit is restricted to magnitudes brighter than  M$_{\rm{UV}} =-15$ (marked by the vertical line). For comparison, $z \sim 6$ UV LF results from B17, L17, and I18 are also shown with the red, green, and brown curves, respectively. For consistency, following B17, we shift the B17 and L17 LFs down by 0.15 dex, corresponding to the expected redshift evolution of the LF between their z=6 sample and our combined z=6-7 sample.  {\bf Right:} same as left panel but measurements are now fitted with a modified Schechter function (cf. Eq. \ref{eq:turn}) that allows a curvature fainter than M$_{\rm{UV}} =-16$. Note that among the literature results only \citet{bouwens17b} allow a turnover at faint magnitudes.}
   \label{fig:uv_lf}
\end{figure*}

We now compute the final UV luminosity function at z$\sim$6 resulting from the combination of all the individual clusters. In order to account for all the uncertainties described in the previous sections, we use Markov Chain Monte Carlo (MCMC) simulations to explore the full error space of each galaxy and each effective volume in a given magnitude bin. At the level of an individual galaxy, the MCMC simulations take into account the photometric scatter and magnification factor uncertainties by using multiple lensing models. Then, we compute the effective survey volume $V_{eff}$ for each galaxy by randomly sampling the completeness functions (and lensing models) shown in the left side of Fig. \ref{fig:models} for the corresponding cluster. For each iteration, the total sample is distributed in magnitude bins $M_{i}$, with a size of 0.5 mag, for which $\phi(M_{i})$ is computed with the associated cosmic variance errors estimated by \citet{robertson13}. These simulations have the advantage of accounting for uncertainties in $M_{i}$ when estimating uncertainties on $\phi(M_{i})$ because individual galaxies can change bins at each iteration, and at the same time, rectifying the limitations of the binning procedure. Simultaneously, we also fit the resulting LF results at each iteration to determine the likelihood of the modified Schechter parameters. We here combine our results with the bright-end constraints derived by \citet{bouwens15} from wide-area blank field surveys.

The final LF results are shown in Figure \ref{fig:uv_lf}, together with the LF values listed in Table \ref{tab:lf}. We show on the left panel the best-fit constraints using a classical Schechter function with the associated confidence regions. It is noteworthy that, unlike LF measurements that extend to fainter magnitudes with a nearly constant slope (I17,L17), there is clear drop in our LF at faint magnitudes. Therefore, it is sensible to truncate the luminosity range to bins brighter than M$_{\rm{UV}} =-15$  when using a Schechter function with a constant faint-end slope. We find a faint-end slope of $\alpha = - 1.98_{-0.09}^{+0.11}$,which is steeper than the B17 estimate of $\alpha = -1.91 \pm 0.02$, but shallower than estimates of L17 and I18 with $\alpha = -2.10_{-0.03}^{+0.03}$ and $\alpha = -2.15^{+0.08}_{-0.06}$, respectively. Importantly, the uncertainties of the best-fit parameters derived in the present work are significantly larger than those studies.

A better comparison is achieved with the best-fit function that allows for a curvature at very faint magnitudes following Eq. \ref{eq:turn}. Now we obtain a best-fit faint-end slope of $\alpha = -2.01_{-0.14}^{+0.12}$ and a curvature parameter of $\beta =  0.48_{-0.25}^{+0.49}$. Our results show a stronger turnover than what is found by B17, who extend the LF to $\sim -13$ mag with a slight turnover of $\beta =  0.17 \pm 0.2$. From equation \ref{eq:turn}, we determine the turnover magnitude of the LF at $M_{T} = -14.93_{-0.52}^{+0.61}$, which is close to the result of B17 at $M_{T} = -14.9$. Using empirical results from four HFF clusters and a luminosity function model, \citet{yue17} define $M_{T}$ as the magnitude at which the the classical Schechter LF drops by a factor of 2 and report $M_{T} \geq -14.3$. Comparing to our definition and their Figure 5, they find a similar turnover magnitude but with a shallower curvature. The results of L17 show a strong evidence against a turnover at magnitudes brighter than M$_{\rm{UV}} =-12.5$ at $z=6$.

The discrepancies observed between our results and other studies could partially be explained by the larger number of galaxies reported in the sample of L17 due to a less drastic selection criteria and a more sophisticated correction for ICL contamination, or by a larger size distribution in their completeness simulations (cf. Secion \ref{sec:err-size}). B17 find a much shallower curvature in the LF at a similar magnitude compared to our study. They use very small sizes (around 3 mas) for faint galaxies in their completeness simulations based on the latest results of high-z studies including the HFF clusters. Such distribution naturally leads to a shallower slope at the faint-end. Most importantly, the uncertainties derived from our end-to-end simulations prevent us from putting robust constraints on the very faint-end shape of the LF.  Most of the lensing and sizes uncertainties investigated by this procedure impact the faint-end of the LF for which the 2$-\sigma$ confidence region allows a wide range of curvature and slope parameters.

\begin{table}
\centering
\caption{The UV LF determination at $z \sim 6$ as shown in Figure \ref{fig:uv_lf}.}
\label{tab:lf}
\begin{tabular}{l c }
\hline
\hline
  M$_{\rm{UV}}$ & $\log_{10} \phi(M)$  \\
   AB mag & Mpc$^{-3}$ mag$^{-1}$      \\
\hline
 -21.86 &    $-4.58  \pm   0.15  $   \\
 -21.36 &    $-4.57    \pm   0.19   $  \\
 -20.86 &    $-4.03     \pm  0.12    $ \\
 -20.36 &    $-3.64    \pm  0.09    $ \\
 -19.86 &    $-3.32   \pm   0.08 $    \\
 -19.36 &    $-3.05    \pm  0.08  $   \\
 -18.61 &    $-2.69   \pm   0.08   $  \\
 -18.25 &    $-2.42   \pm   0.09  $   \\
 -17.75 &    $-2.22    \pm  0.08   $  \\
 -17.25 &    $-1.98   \pm   0.08  $   \\
 -16.75 &    $-1.81    \pm   0.20  $   \\
 -16.25 &    $-1.71    \pm   0.15 $    \\
 -15.75 &    $-1.54    \pm   0.36  $   \\
 -15.25 &    $-1.36   \pm    0.37  $   \\
 -14.75 &    $-1.39   \pm    0.71  $   \\
 -14.25 &    $-1.38    \pm   0.53 $    \\
 -13.75 &    $-1.47    \pm   0.73 $    \\ 
 -13.25 &    $-1.56     \pm   1.40  $   \\
 \hline

\end{tabular}
\end{table}

\begin{table*}
\centering
\caption{Best fit constraints on the $z \sim 6$ UV luminosity function}
\label{tab:lf}
\begin{tabular}[c]{l c c c c c}
\hline
\hline
Reference &  $M_{UV}^\star$ & $\alpha$ & $\log_{10} \phi^\star$  & $\beta$ $^{c}$   & $M_{T}$ $^{d}$ \\
 &  [AB mag]& & [Mpc$^{-3}$] & & [AB mag] \\
\hline
This work  $^{a}$           	& $-20.74_{-0.20}^{+0.21}$	& $-1.98_{-0.09}^{+0.11}$	& $-3.43_{-0.21}^{+0.21}$	  & --  & -- \\
\citet{atek15b} 	& $-20.90^{+0.90}_{-0.73}$	& $-2.01^{+0.20}_{-0.28}$	& $-3.55_{-0.57}^{+0.57}$	 &-- & --\\
\citet{bouwens17b}$^{e}$ 	& $-20.94$	& $-1.91 \pm 0.02$	& $-3.18 \pm 0.03$	   &  --  &-- \\ 
\citet{ishigaki18} & $-20.89^{+0.17}_{-0.13}$	& $-2.15^{+0.08}_{-0.06}$	& $-3.78^{+0.15}_{-0.15}$	 &-- & -- \\
\citet{livermore17}	& $-20.82^{+0.04}_{-0.03}$	& $-2.10_{-0.03}^{+0.03}$	& $-3.64^{+0.04}_{-0.03}$	 & -- & --\\
\hline

This work  $^{b}$           	& $-20.84_{-0.30}^{+0.27}$	& $-2.01_{-0.14}^{+0.12}$	& $-3.54_{-0.07}^{+0.06}$	 & $0.48_{-0.25}^{+0.49}$& $-14.93_{-0.52}^{+0.61}$ \\
\citet{bouwens17b}$^{e}$	& $-20.94$	& $-1.91 \pm 0.04$	& $-3.24 \pm 0.04$	 & $ 0.17 \pm 0.2 $& -14.9  \\
\hline
\multicolumn{6}{l}{ $^\textrm{a}$ Using a Schechter functon fit} \\
\multicolumn{6}{l}{ $^\textrm{b}$ Allowing for a turnover in the LF fainter than M$_{\rm{UV}} = -16$ mag} \\
\multicolumn{6}{l}{ $^\textrm{c}$ The turnover parameter presented in equation \ref{eq:turn} that quantifies the curvature} \\
\multicolumn{6}{l}{ of the LF at magnitudes fainter than M$_{\rm{UV}} = -16$ mag}\\
\multicolumn{6}{l}{ $^\textrm{d}$ The turnover magnitude at which the LF departs from the simple Schechter function} \\
\multicolumn{6}{l}{ $^\textrm{e}$  Using the CATS model }
\end{tabular}
\end{table*}

\begin{figure}
   \centering
   \includegraphics[width=8.8cm]{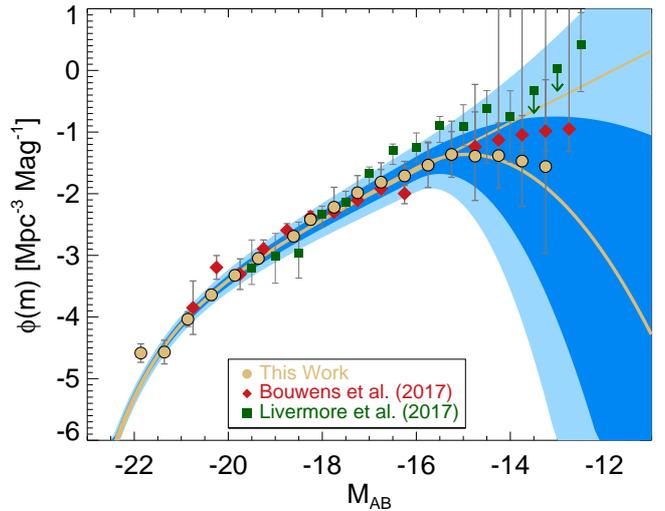}  
   \caption{The final UV luminosity function at $z \sim 6$ compared to recent determinations in the literature. The two best-fit curves are shown in gold while LF measurements of \citet{livermore17} and \citet{bouwens17b} are shown with green squares and red diamonds, respectively. These two literature curves were shifted down by 0.15 dex to account for the difference in the mean redshift of the LF described in the text.}
   \label{fig:uv_lf_points}
\end{figure}

\section{The $z \sim 6$ UV Luminosity Density}
\label{sec:density}

\begin{figure}
   \centering
   \includegraphics[width=8.8cm]{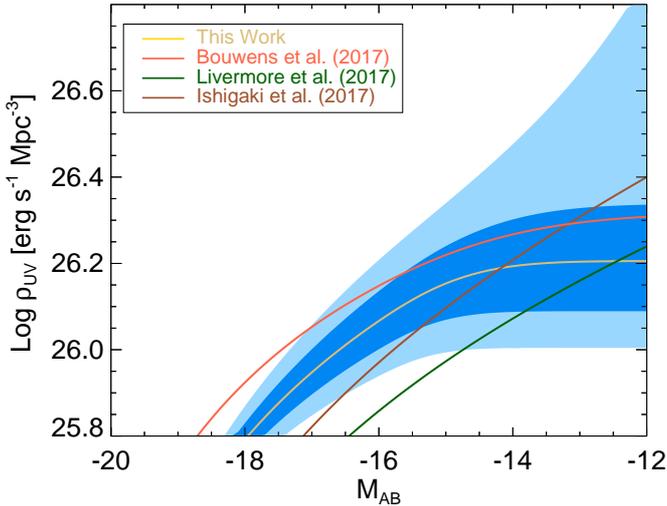}  
   \caption{The UV luminosity density at $z \sim 6$ as a function of the lower integration limit of the UV LF is illustrated by the gold curve.  The shaded light (dark) blue regions are the 1-$\sigma$ (1-$\sigma$) uncertainties. Other $\rho_{\rm UV}$ determinations of B17, L17, and I18 are shown with the red, green, and brown curves, respectively. Note that the LF of B17 and L17 used to compute $\rho_{\rm{UV}}$ was shifted down by 0.15 dex to account for a redshift evolution (cf. Section \ref{sec:errors})}
   \label{fig:rho_uv}
\end{figure}

Using our UV LF determination we now compute the total galaxy UV luminosity density $\rho_{\rm UV}$ at $z = 6$. The MCMC simulations performed for the constraints on the LF parameters are propagated to compute the confidence intervals of the luminosity density. We investigate different truncation magnitudes (the faint integration limit of the LF) to determine to what extent galaxies can produce enough UV radiation to sustain reionisation. The UV luminosity density as a function of the limiting magnitude is shown in Fig. \ref{fig:rho_uv}. Beyond a magnitude of M$_{\rm{UV}} =-15$, which corresponds to the turnover of the luminosity function, the luminosity density becomes nearly flat (on a log scale), reaching Log($\rho_{\rm UV}$/erg s$^{-1}$ Mpc$^{-3}$) $= 26.15 \pm  0.09$ at M$_{\rm lim} =-15$ and Log($\rho_{\rm UV}$/erg s$^{-1}$ Mpc$^{-3}$) $= 26.21 \pm  0.13$ at M$_{\rm lim} =-10$. In comparison, B17 report a luminosity density of Log($\rho_{\rm UV}$/erg s$^{-1}$ Mpc$^{-3}$) $= 26.33$ at M$_{\rm lim} =-15$, slightly larger than our determination due to their larger value of $\phi^{\star} = 0.58 \times 10^{-3}$ Mpc$^{-3}$. In the case where a Schechter function with a constant faint-end slope is used, I18 and L17 find smaller values resulting from different constraints used in the bright-end of the LF, which yield smaller values for $\phi^{\star}$. When we use a Schechter fit, we find Log($\rho_{\rm UV}$/erg s$^{-1}$ Mpc$^{-3}$) $= 26.41 \pm 0.14$ at M$_{\rm lim} =-10$
Again, the confidence region can accommodate a wide range of values for the total luminosity density and the shape of its evolution with the truncation magnitude and remains compatible with most of the literature results. Therefore, it is clear that accounting for te lens model uncetainties no robust constraints can be inferred regarding the existence of a large population of faint (M$_{\rm UV} > -15$) galaxies that could provide the required energy budget to reionize the Universe.

\section{Conclusion}
\label{sec:conclusion}
The lensing clusters of the Hubble Frontier Fields recently made it possible to explore the distant Universe to the faintest magnitude limits ever achieved. Based on the unprecedented quality of the {\em HST} images, several groups have constructed mass models in order to interpret high-redshift observations. For instance, the galaxy luminosity function and the ability of faint galaxies to reionize the Universe was the focus of multiple studies from different groups \citep[e.g.,][]{atek15b,bouwens17b,livermore17,ishigaki18,yue17}. With the completion of the HFF program, significant differences were found between publications at the very-faint-end of the LF, where the lensing magnification, and their associated complex uncertainties become important. 

In the present paper, we used the six HFF clusters data to construct a robust UV LF at $z \sim 6$. We have used for the first time a comprehensive approach based on end-to-end simulations that includes all lensing uncertainties and their impact on the study of the UV luminosity function at $z \sim 6$. Our approach uses simulated galaxies directly in the source plane, which has the advantage of accounting for all lensing uncertainties present in both effective survey volume and the number counts.       

We first used this procedure to assess the impact of the source size distribution on the survey volume, hence on the UV LF at $z \sim 6$. We find that uncertainties on the size-luminosity relation has a significant impact on the faint-end of the UV LF. Adopting large sizes with a half light radius hlr = 50 mas leads to a small recovery fraction therefore to a steeper faint-end slope \citep[e.g.,][]{livermore17}. Smaller sizes with half light radii below 10 mas lead to a shallower slope \citep{bouwens17b}. Overall, such uncertainties create a wide range of slopes beyond $M_{\rm UV} =-15$ mag.

Our procedure has the second advantage of using any mass model and therefore provides means to assess the systematic lensing uncertainties. We achieve this goal by comparing the results using four different models to compute the effective survey volume and the UV LF. We show that the combination of systematic and intrinsic uncertainties lead to important differences in the final UV luminosity function beyond an intrinsic magnitude of $M_{\rm UV} =-15$, where most galaxies will have magnification factor greater than 10. 

Finally, we computed the UV luminosity function while incorporating the different uncertainties discussed in this paper. Adopting a simple Schechter fit we find a faint-end slope of $\alpha = - 1.98_{-0.09}^{+0.11}$, whereas a modified Schechter function that permits curvature in the LF at M$_{\rm UV} > -16$ mag yields a turnover in the LF with a faint-end slope of $\alpha = - 2.01_{-0.14}^{+0.12}$ and a curvature parameter of $\beta =  0.48_{-0.25}^{+0.49}$. Most importantly, while galaxies were detected down to an intrinsic magnitude of M$_{\rm UV} \sim -13$, we were unable to reliably extend the UV LF beyond M$_{\rm UV} \sim -15$ because of the large confidence interval. Consequently, the existence of a large reservoir of faint galaxies that significantly contribute to the total UV luminosity density is still uncertain.

Overall, we demonstrated that with the current depth of observations and current state-of-the art in mass modeling of lensing clusters, robust constraints on the UV luminosity function fainter than $M_{\rm UV} =-15$ mag remain unrealistic. Future observations of lensing clusters with the upcoming {\em James Webb Space Telescope} will push observed flux limits by about two magnitudes and at the same time provide hundreds of spectroscopic redshifts of multiple images to improve the accuracy of lensing models. Such observations will therefore bring a definitive answer to the potential turnover in the UV LF and the contribution of extremely faint galaxies to cosmic reionization.

\section*{Acknowledgements}

We want to thank the STScI and the HFF team for their efforts in obtaining and reducing the {\em HST} data. HA is supported by the Centre National d'Etudes Spatiales (CNES). JPK is supported by the European Research Council (ERC) advanced grant ``Light in the Dark'' (LIDA). JR acknowledges support from the ERC starting grant CALENDS. 
\bibliographystyle{mnras}
\bibliography{references}

\bsp	
\label{lastpage}
\end{document}